\documentclass{emulateapj}

\def\ga{\mathrel{\hbox{\rlap{\hbox{\lower4pt\hbox{$\sim$}}}\hbox{$>$}}}}
\def\la{\mathrel{\hbox{\rlap{\hbox{\lower4pt\hbox{$\sim$}}}\hbox{$<$}}}}

\usepackage{verbatim}
\usepackage{amsmath}
\usepackage{multirow}
\usepackage{relsize}
\usepackage{natbib}
\usepackage{color}

\usepackage{amssymb}
\usepackage{float}

\shorttitle{Radio Auroral Emission from Proxima B}

\shortauthors{Burkhart and Loeb}
\begin{document}
\title{The Detectability of Radio Auroral Emission from Proxima B}
\author{Blakesley Burkhart \& Abraham Loeb\altaffilmark{1} }
\altaffiltext{1}{Harvard-Smithsonian Center for Astrophysics, 60 Garden st. Cambridge, Ma, USA}

\begin{abstract}
Magnetically active stars possess stellar winds whose interaction with planetary magnetic fields produces radio auroral emission.  We examine the detectability of radio auroral emission from Proxima b, the closest known exosolar planet orbiting our nearest neighboring star, Proxima Centauri.  Using the Radiometric Bode's Law, we estimate the radio flux produced by the interaction of Proxima Centauri's stellar wind and Proxima b's magnetosphere for different planetary magnetic field strengths. For plausible planetary masses, Proxima b produces 6-83 mJy of auroral radio flux at frequencies of 0.3-0.8 MHz for  planetary magnetic field strengths of 1-3 B$_{\oplus}$. According to recent MHD models that vary the orbital parameters of the system, this emission is expected to be highly variable. This variability is due to large fluctuations in the size of Proxima b's magnetosphere as it crosses the equatorial streamer regions of the dense stellar wind and high dynamic pressure.
Using the MHD model of \citet{Garraffo2016} for the variation of the magnetosphere radius during the orbit, we estimate that the observed radio flux can vary nearly by an order of magnitude over the 11.2 day period of Proxima b. The detailed amplitude variation depends on the stellar wind, orbital, and planetary magnetic field parameters. We discuss observing strategies for proposed future space-based observatories to reach frequencies below the ionospheric cut off ($\sim 10$ MHz) as would be required to detect the signal we investigate.

\end{abstract}


\section{Introduction}
\label{intro}

All planets with substantial magnetic fields in our Solar System strongly emit coherent low-frequency radio cyclotron emission associated with auroral activity. This radio emission is produced by active auroral regions with electrons propagating around magnetic field lines \citep{Gurnett1974,Zarka1992}
and is proportional to the intercepted magnetic energy flux around the radius of the planetary magnetosphere. Although radio emission from planetary aurora is well-studied in the Solar System, it has yet to be detected for exoplanetary systems \citep{Griessmeier2007,Luger2017}. 
Detecting radio emission from exoplanets would allow for further constraints on their orbital parameters \citep{Luger2017}, open up new detection methods at radio wavelengths \citep{Farrell1999}, and allow us to measure exoplanet magnetospheres for the first time \citep{Shkolnik2008,Vidotto2010,Kislyakova2014}. Furthermore, measurements of magnetic fields via auroral emission would also shed light on exoplanet habitability and atmospheric composition \citep{Lundin2007,Griessmeier2016,Luger2017}.

The discovery of the exosolar planet Proxima b orbiting in the potential “habitable zone” of our
nearest stellar neighbor Proxima Centauri \citep{Anglada2016} presents the closest (1.3 parsec away) candidate for possibly detecting an exo-aurora. 
Proxima b is estimated to be at least 1.3 Earth masses (with an inclination-corrected mass of Msin($i$) = 1.3M$_{\oplus}$) and has an orbital
period of 11.2 days. Its orbit has a semi-major axis of 0.049 AU; twenty times closer to Proxima Centauri than
the Earth is to the Sun \citep{Anglada2016}.
Due to its 1.3 pc distance from our Solar System,  Proxima b could be directly observable by the next generation of space telescopes
such as ELT, WFIRST and JWST \citep{Barnes2016,Kreidberg2016,Luger2017}  and has been the subject  of studies examining its likely irradiation history, possible climate, and evolution in relation to potential habitability \citep{Turbet2016,Barnes2016,Meadows2016,Ribas2017}.

\citet{Garraffo2016} studied the space weather of Proxima b using numerical MHD simulations of the stellar wind from Proxima Centauri. The authors examined the wind conditions (densities, pressures and magnetic field strength) over different plausible orbits of Proxima b. They found that the planet is subject to stellar wind pressures of more than 2000 times those experienced by Earth from the solar wind. Magnetic activity can be enhanced due to interactions between close-in planets and their host stars, similar to the interaction of the Io-Jupiter system \citep{Crary1997}. Based on MHD models and Solar System examples,  it is therefore likely that, if Proxima b has an Earth-like magnetic field, auroral emission could be detectable from low frequency space-based radio arrays such as the proposed Radio Observatory on the Lunar Surface for Solar studies (ROLSS) \citep{Lazio2011,Zarka2012}.
Future ground based telescopes, such as an Extremely Large Telescope (ELT), could possibly observe the exo-aurora of Proxima b using optical emission lines \citep{Luger2017}.

In this paper we examine the detectability of radio auroral emission from Proxima b.  Using the radiometric Bode's Law, we estimate the expected radio flux and frequency emitted from  Proxima b and compare that to previous estimates for other exoplants in Section \ref{sec:scaling}. In Section \ref{sec:rm_rp} we calculate the expected radio light curves for Proxima b based on the space weather modeling of \citet{Garraffo2016} and for varying orbital parameters. Finally, in Section \ref{sec:disc} we discuss our results, followed by a summary of our conclusions in Section \ref{sec:con}.

\section{The Radiometric Bode's Law: Stellar-Wind Driven Cyclotron Emission}
\label{sec:scaling}
First, we address the expected order-of-magnitude radio flux density from low frequency radio observations of the Proxima b system.
To estimate the radio power being emitted from the interaction of the stellar-wind and Proxima b's magnetosphere we use the Radiometric Bode's law for the solar system, where the planetary radio power, $P_{\rm{radio}}$, can be estimated from the power released by the dissipation of stellar wind kinetic or magnetic energy \citep{Desch1984}.
Several variants on Bode's Law exist in the literature and provide an \textit{empirical} scaling relation for observed radio power to planetary quantities such as planetary mass ($m$), magnetic moment ($\cal{M}$), rotation speed ($\omega$), planetary radius ($r$), and planetary distance from the host star ($d$).

While the Radiometric Bode's Law is an empirical result derived for the Solar System, its application for detecting radio emission from exoplanets has been discussed extensively \citep{Desch1984,Zarka1992,Farrell1999,Farrell2004,Lazio2004}.
The radio auroral power for a planetary magnetosphere increases in proportion to the size of the magnetosphere cross-section, which depends on the planetary magnetic moment ($\cal{M}$). 
 Assuming the magnetic moment scales with planetary mass and rotation speed, e.g. via the so-called Blackett's law \citep{Blackett1947}, $\cal{M}$ $\propto \omega m^{5/3}$, a radiometric Bode's law based on Solar System scalings proposed by \citet{Zarka1992} can be written as:

\begin{equation}
P_{\rm{radio}}\approx 4\times 10^{11}\left(\frac{\omega}{\omega_J}\right)^{0.79} \left(\frac{m}{m_J}\right)^{1.33}\left(\frac{d}{d_J}\right)^{-1.6} \rm{W}
\label{eq:zarka}
\end{equation}
with all planetary qualities normalized to Jupiter values. 

We adopt fiducial values for Proxima b of $d=0.049$ AU and a mass of m=2 M$_{\oplus}$.  Proxima b is most likely tidally locked due to its close-in orbit, and therefore we assume the rotation period is the same as the orbital period of $\omega=11.2$ days.
Using these values, Equation (\ref{eq:zarka}) gives a predicted radio power for Proxima b
of $P_{\rm{radio,prox}}=8\times 10^{11}$W.
For comparison, Earth's observed and predicted radio power using Equation (\ref{eq:zarka}) is two orders of magnitude lower than Jupiter and Proxima b.

In order to calculate the observed radio flux at Earth we assume Proxima b has an Earth-like magnetic field range of B$_{\rm{prox}}$=0.1-0.3 G. This gives a cyclotron frequency,
$f=eB/2\pi m_e$=0.28-0.84 Mhz. 

The radio flux density is given by:
\begin{equation}
\Phi=\frac{P_{\rm{radio}}}{4\pi d^2 \delta f} \frac{\rm W}{\rm m^2 Hz}
\end{equation}
assuming an emission cone of $4\pi$ sr and $\delta f=f/2$.
Plugging in the fiducial values for Proxima b yields a flux density of $\Phi_{\rm{proxima}}=6-19$ mJy for magnetic fields of B$_{\rm{prox}}$=0.3-0.1 Gauss, respectively. We note that this value can increase if the mass of Proxima is larger. For example using B$_{\rm{prox}}=0.1$G and $m=6M_{\oplus}$, yields $\Phi_{\rm{proxima}}=83$ mJy.

\begin{figure}
\includegraphics[width=9.8cm]{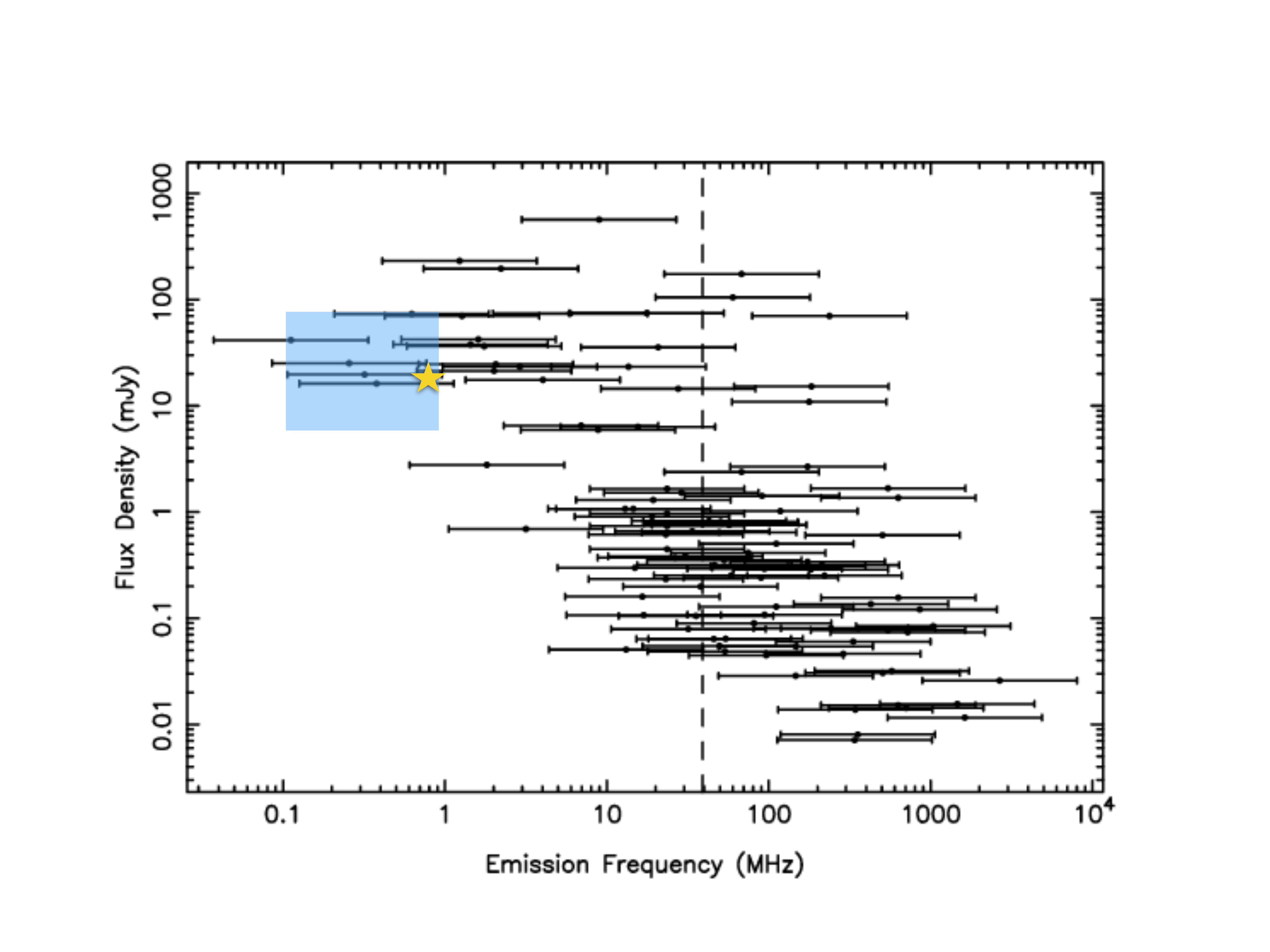}
\caption{
\label{fig:proximaexo}
Predicted flux densities vs. characteristic emission frequency based on the Radiometric Bode's law and Blackett's law for Proxima b. The expected range discussed in the text is marked with a blue box and yellow star. Overplotted are radio fluxes for 106 exoplanets (black lines, described in Lazio et al. 2004), also calculated using Equation (\ref{eq:zarka}). As a reference, the vertical dashed line indicates the cut-off frequency for Jupiter.  Below $\sim$10 MHz, ground-based observations are blocked by the ionosphere. Modified from \citet{Lazio2004}
}
\end{figure}

We summarize these calculations in Figure \ref{fig:proximaexo} with a comparison of the values
of radio power and observing frequency obtained for Proxima b (bracket in the blue box) to
106 exosolar planets (black lines, taken from Lazio et al. 2004), all calculated using Equation (\ref{eq:zarka}).  Given the uncertainties in the values of magnetic field and mass of Proxima b, we bracket in blue the range of values below the fiducial 19 mJy value calculated for $m=2 M_{\oplus}$ and $B_{\rm{prox}}=0.1$G (marked with a yellow star).  Due to its Earth-like size and extremely close proximity to its magnetically active host star, Proxima b should show a strong radio signature. However, below $\approx$10 Mhz ground observations are blocked by the ionosphere and therefore space-based low frequency observations will be required. We address this point further in the discussion section.

\section{Orbital Variability of Radio Emission}
\label{sec:rm_rp}
While our estimates for the radio power of Proxima b in Section \ref{sec:scaling} provide order of magnitude estimates, they do not account for the variability of the magnetosphere during the 11.2 day orbit. Proxima Centauri is an M-dwarf star which is expected to have a high mass loss rate, with an upper limit of $\dot{M}\approx3\times10^{-13} M_{\odot}yr^{-1}$ \citep{Wargelin2002}, a highly variable stellar wind, and frequent flaring events \citep{Hawley2014}.
Therefore the variation in the possible observed radio flux of Proxima b could be large and be an important observational signature.

Using the BATS-R-US MHD code \citep{Holst2014}, \citet{Garraffo2016} recently constructed 3D MHD models of the wind and magnetic field around Proxima Centauri b using a surface magnetic field map for a star of the same spectral type. They scaled this model to match the expected average $\sim$600 G surface magnetic field strength of Proxima Centauri.
They found that Proxima b's magnetopuase standoff distance is:
\begin{equation}
r_m=\left(\frac{B_P^2}{4\pi P_{W}}\right)^{1/6}R_p,
\label{eq:rm}
\end{equation}
where $P_{W}$ is the ram pressure of the stellar wind, $R_p$ is the planetary radius ($R_{p,prox.}=1.1R_{\oplus}$, and B$_{P}$ is the planetary magnetic field strength, undergoes sudden and periodic changes by a factor of 2–5 and is bombarded with a solar wind pressure and density of order 10$^3$ times that of Earth.  The exact amplitude change depends on the choice of the input  $B_P$ as well as the orbital inclination, $i$, between the orbit of the planet and the rotational axis of the star. These rapid changes in $r_m$ occur twice each orbit as the planet passes through the equatorial streamer
regions of the dense wind and high dynamic pressure on a timescale as short as a day (see e.g. Figure 5 of Garraffo et al.2016).
We therefore adopt $r_m$ fluctuations during Proxima b's orbit to make a more physically realistic assessment of the auroral radio flux.

\begin{figure}
\includegraphics[width=9.6cm]{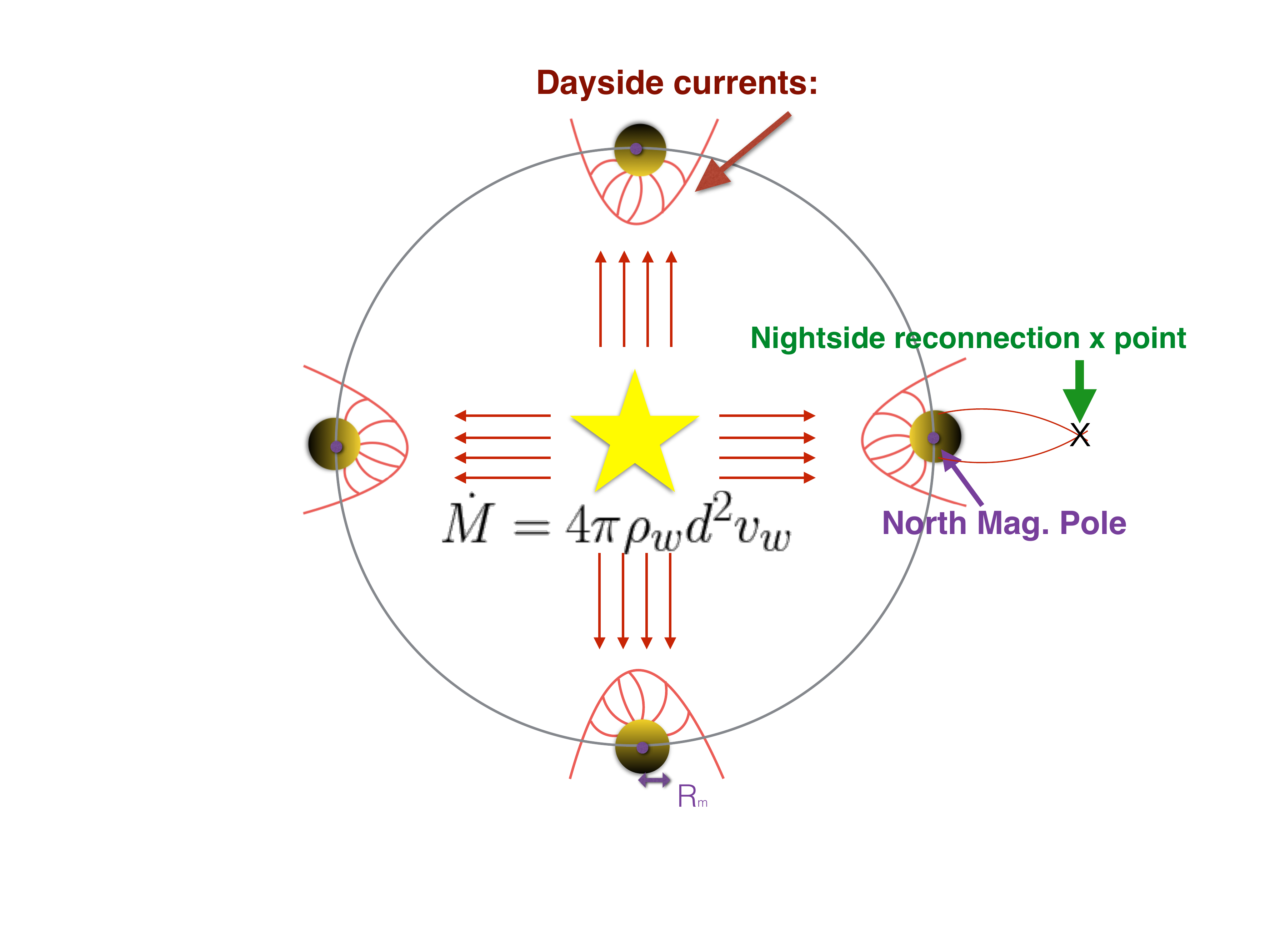}
\caption{Illustration of the active regions in a typical planetary magnetosphere and their associated auroral
regions. Dayside currents deposit energy in the upper atmosphere mediated by the planetary magnetic field. 
These currents create the aurora and subsequent radio emission discussed here. Another potential source of current is at the nightside reconnection x point, where fields of opposite polarity converge.  Nightside reconnection emission is not considered here.
\label{fig:cartoon}}
\end{figure}

Due to the high dynamic pressure of the stellar-wind and relatively weak planetary magnetic field, the net dissipated power on the planet day side of Proxima b is expected to be larger than in the reconnection region of the magnetotail on the planet's night
side, by more than an order-of-magnitude \citep{Varela2016}.
For exoplanets that experience a stellar-wind of lower dynamic pressure and have more intense magnetic fields, the configuration can be the opposite, leading to stronger radio emission from the reconnection region of the magnetotail and weaker radio emission from the planet's dayside.
For the purposes of the estimations in this paper, we will consider that most of the radio emission is coming from the dayside of Proxima b.  Figure \ref{fig:cartoon} shows the basic schematic view of the system.

The dissipated magnetic and kinetic powers of the impacting
wind on the planet are approximated as:
\begin{equation}
P_B=\frac{B_{w}^2 \pi r_m^2 v_{w}}{4\pi}
\label{eq:pb}
\end{equation}
\begin{equation}
P_k=\rho_w v_{w}^3 \pi r_m^2
\label{eq:pk}
\end{equation}
where $v_{w}$ is the 
relative velocity between the wind
and the Keplerian velocity of the planet, $\rho_w$ is 
the local wind density, and $B_{w}$ is the interplanetary magnetic field component perpedndicular to the solar wind flow  impinging on the planet's magnetospherehe  \citep{Zarka2007,Vidotto2017}. 

In order to estimate the radio power being emitted from the interaction of the stellar wind and Proxima b's magnetosphere, we adopt the model of \citet{Garraffo2016} for the variation of $r_m$ as a function of orbital period and apply again the radiometric Bode's law for the solar system as given in \citet{Vidotto2010,Vidotto2017}. The planetary auroral radio power ($P_{radio}$) can be estimated from the power released by the dissipation of stellar wind kinetic power, $P_k$, and/or the power released from the dissipation of magnetic power, $P_B$, of the wind \citep{Farrell1999,Zarka2001,Zarka2007,Vidotto2010}. We consider both here for completeness; however we expect that the magnetic power is likely to be
dominant for converting flow power to energetic particles \citep{Zarka2010,Vidotto2017}.

The radiometric Bode's law is then given as:
\begin{equation}
P_{\rm{radio,k}}=\nu_k P_k,
\label{eq:pk2}
\end{equation}
and 
\begin{equation}
P_{\rm{radio,B}}=\nu_B P_B,
\label{eq:pb2}
\end{equation}

where $\nu_k$ and $\nu_B$ are efficiency ratios whose values we adopt from  the
Solar System data \citep{Zarka2007,Vidotto2017}, $\nu_k=1\times10^{-5}$ and $\nu_B=2\times10^{-3}$. \footnote{ We stress that these efficiency coefficients are based on data from our Solar System and could be different for M-dwarf stars. However the overall estimated mass loss rate from Proxima Centauri is estimated to be no more than an order of magnitude higher than the sun \citep{Wood2001}.} 

Based on the strong field model of \citet{Garraffo2016}, we assume $v_{w}=1600  \rm{kms^{-1}}$, $\rho_w/m_p=1000 \rm{cm^{-3}}$ and B$_{w}=0.01$G (representative of a 1200 Gauss dipole field at the surface of a star with radius of 0.2R$_{\bigodot}$ at a distance of Proxima b).
In agreement with \citet{Garraffo2016}, our parameters produce a wind dynamic pressure in excess of the wind magnetic pressure by about an order-of-magnitude.   We investigate three possible values of Proxima b's polar magnetic field: $B_p$=0.1,0.3,1G and scale the values of $r_m$ vs. orbital phase given in \citet{Garraffo2016} according to Equation (\ref{eq:rm}). We then apply Equations (\ref{eq:pb}-\ref{eq:pb2}) and 
calculate the expected radio fluxes as viewed from Earth.

\begin{figure}
\includegraphics[width=9.8cm]{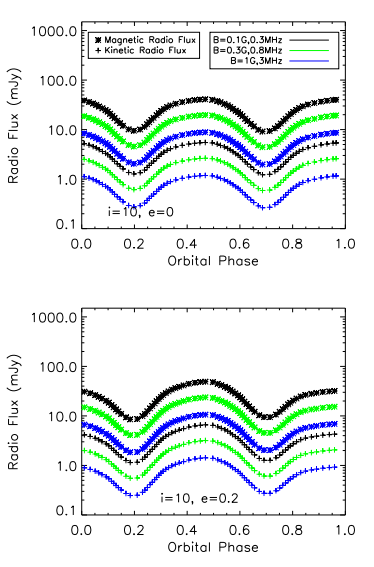}
\caption{Variation in the observed radio flux as the magnetospheric radius, $r_m$, varies over the planetary orbital period of 11.2 days. We use the MHD model of \citet{Garraffo2016} for the magnetospheric radius and apply Equations (\ref{eq:pk}) and (\ref{eq:pb}) to calculate the expected kinetic (plus symbols) and magnetic (asterisks symbols) radio power and flux. Here we consider a high orbit between the planet and rotation axis of the star of i=60$^o$.  The top panel has orbital eccentricity of e=0 and the bottom panel has e=0.2.
\label{fig:i60}}
\end{figure}

\begin{figure}
\includegraphics[width=9.8cm]{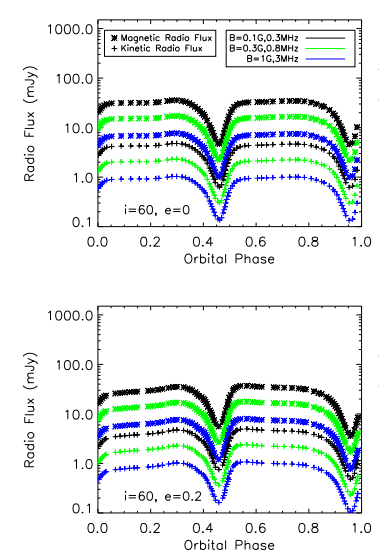}

\caption{Same as Figure \ref{fig:i60} except for a low orbit between the planet and rotation axis of the star of i=10$^o$.
\label{fig:i10}}
\end{figure}

Figures \ref{fig:i60} and \ref{fig:i10} show the variation in the observed radio flux vs. orbital phase for orbital inclination, $i$, between the orbit of the planet and the rotational axis of the star. The top panels show a planetary eccentricity of $e=0$  and the bottom panels show $e=0.2$. While there is a dependency on orbital eccentricity, the dominant dependency of the light curve on orbital parameters is on the inclination between the orbit and the stellar rotation axis. Figure \ref{fig:i10} adopts $i=10$ while Figure \ref{fig:i60} assumes $i=60$.  
Magnetic power (Equation \ref{eq:pb}) is shown with asterisks symbols while kinetic power (Equation \ref{eq:pk}) is shown using plus symbols. The overall radio flux output from the magnetic power tends to dominate the output from kinetic power. Different line colors indicate varying planetary magnetic field strength. 
Depending on the orbital inclination, Proxima b's aurora can vary by nearly an order of magnitude over the 11.2 day period. The overall values of radio flux are consistent with the range reported in the previous section.


\section{Discussion}
\label{sec:disc}

\begin{figure}
\includegraphics[width=8.5cm]{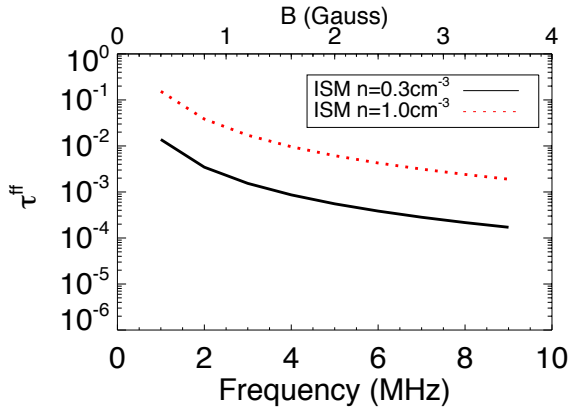}
\caption{
\label{fig:freefree}
Optical depth for free-free absorption ($\tau_{ff}$) from the local interstellar medium along the line of sight to Alpha Centauri as a function of observing frequency. Lower frequency emission (corresponding to lower values of magnetospheric magnetic field strength) is more likely to be absorbed (i.e. reach a value of unity on the y axis). Due to the proximity of the Alpha Centauri system, free free absorption is not problematic, however for more distant planetary systems low frequency observations will be challenging.
}
\end{figure}

In order to detect radio cyclotron emission from any exoplant we must be above the threshold for free-free absorption from electrons in the local ionized interstellar medium (ISM). The absorption coefficient as given in Equation (5.19b) of \citet{Rybicki1979} is:
\begin{equation}
\alpha_{\nu}^{ff}\approx0.018 T^{-3/2}\nu^{-2}n^2 cm^{-1}
\end{equation}
where $n$ is the interstellar density in cm$^{-3}$, $T$ is the temperature in K, and $\nu$ is the frequency of the radiation in Hz.
Figure \ref{fig:freefree} shows the free-free absorption from the local interstellar medium integrated along the line of sight to Alpha Centauri (1.3 parsec) as a function of observing frequency for two different values of ISM density and T=7000K. Lower frequency emission (corresponding to lower values of magnetospheric magnetic field strength) is more likely to be absorbed, owing to the scaling of $\alpha_v^{ff}\propto \nu^{-2}$. Due to the proximity of the Alpha Centauri system, it is unlikely that free-free absorption is problematic. However for more distant planetary systems (e.g. at the distance of the TRAPPIST -1 system near 10 pc) low frequency observations will be absorbed by the ISM.


The Low Frequency Array (LOFAR)\footnote{http://www.lofar.org/} provides the possibility to study weak radio emission from exoplanets with high sensitivity and resolution for 
frequencies between  10 and 240 MHz.  This corresponds to cyclotron emission for magnetic fields down to strengths of 4 G. For lower magnetic field strengths, including the probable field strengths of Earth-like planets such as Proxima b, even arrays operating at  lower frequency will be required. However, Earth's ionosphere significantly impedes any ground based observations below 10 MHz, the plasma frequency cut-off of the ionosphere. 

In order to avoid reflection by the ionosphere, extremely low frequency observations for detection of Earth-like magnetic fields from exoplanets must be done from space.  
 A variety of observatory concepts have been proposed for ultra-low frequency arrays on the lunar surface \citep{Zarka2012}.  One such proposal is the Radio Observatory on the Lunar Surface for Solar studies (ROLSS) \citep{Lazio2011}. An alternate proposal being considered to a low frequency radio array located on the lunar surface is a large array
of CubeSats with dipole electric field
antennas  \citep{Rajan2016}. Our study provides additional motivation for the construction of a space-based low frequency array with very long baselines.  For Proxima b, our study shows that the main objective should not necessarily be to resolve the emission 
but rather to have the sensitivity to observe the variability in the radio flux which would indicate the presence of a varying magnetosphere under the influence of the stellar wind of Proxima Centauri.  Additional scientific goals of a space based low frequency array include studies of the dark ages and epoch of re-ionization via detection of the red-shifted 21cm line \citep{Lazio2011,Fialkov2013,Rajan2016}.

While observations of radio emission for exoplanets below 10 MHz are not feasible with current observatories, auroral signatures could also be detectable via optical emission. 
Towards this goal, \citet{Luger2017} examined the feasibility of detecting auroral emission from Proxima b, would constrain the presence and composition of its atmosphere in addition to determining the planet's
eccentricity and inclination. \citet{Luger2017} searched the Proxima b HARPS data for
any oxygen or nitrogen lines near 5577\AA~ but found no signal, indicating that the OI auroral line contrast must be lower than $2\times10^{−2}$, which was consistent with their model predictions.  Optical studies of Proxima b's possible auroral signatures therefore will also need to wait for future observatories. A space-based coronagraphic telescope or a ground-based Extremely Large Telescope (ELT) with
a coronagraph could push sensitivity down to line contrasts of  $7\times 10^{-6}$ with an exposure on order of 1 day \citep{Luger2017}. 

If aurora are detected around Proxima b, the information gained (constraints on the magnetosphere, orbital inclination and eccentricity) will greatly inform more detailed models and numerical simulations of Proxima Centauri's stellar wind and its interaction with planetary and interstellar magnetic fields \citep{Pogorelov2004}. In situ missions to the Alpha Centauri system may eventually also benefit from auroral detections. The proposed Breakthrough Foundation's Starshot mission\footnote{https://breakthroughinitiatives.org/Initiative/3} aims to launch
gram-scale spacecrafts with miniaturized electronic components
(such as camera, navigation, and communication
systems) to relativistic speeds ($v\approx 0.2c$). This will
enable the spacecraft to reach the nearest stars, like Proxima Centauri, within a human lifetime. Precise orbital measurements, possibly provided by the detection of optical or radio auroral emission, of Proxima b will be important for realizing a close fly-by of this planet that includes images and direct measurements of its magnetic field and atmospheric properties.

\section{Conclusions}
\label{sec:con}
In this work we have used the Radiometric Bode's Law for the Solar System to investigate the possibility of detecting radio auroral emission coming from Proxima b, the closest known exoplanet orbiting our nearest neighboring star, Proxima Centauri.  
We have found that: 
\begin{itemize}

\item Due to its proximity both to Earth and its host star, Proxima b could produce between 4-83 mJy of auroral radio flux at frequencies of 0.3-0.8 MHz assuming a field strength of 1-3B$_{\oplus}$ for a realistic planetary mass range.

\item Based on recent MHD models, the auroral emission should also be highly variable due to the varying size of Proxima b's magnetosphere as it crosses regions of high stellar wind pressure and density.
Using the MHD model of \citet{Garraffo2016} for the variation of the magnetosphere radius during the orbit, we estimate that the observed radio flux can vary an order of magnitude over the 11.2 day period of Proxima b. 

\item The amplitude of the variation depends on orbital parameters as well as the parameters of the stellar wind and planetary magnetic field. This implies that these parameters can be discerned from the detection of auroral variations and detailed MHD modeling.

\item We discuss a number of caveats regarding the detection of radio emission from Proxima b, namely that ultra-low-frequency observatories must be constructed in space in order to avoid blocking by the ionosphere. We show that free-free absorption by the local interstellar medium will not hinder detection.

\end{itemize}

\acknowledgments
 We are grateful for valuable discussions with Dr. Laura Kreidberg, Dr. Cecilia Garraffo and Dr. Zachary Slepian. B.B. acknowledges generous support from the NASA Einstein Postdoctoral Fellowship and the joint Sub-millimeter Array/Institute for Theory and Computation Postdoctoral Fellowship. This work was supported in part by a grant from the Breakthrough Prize Foundation.

\bibliography{refs.bib}

\end{document}